\documentclass[10pt,conference]{IEEEtran}
\usepackage[pdftex]{graphicx}
\usepackage[usenames,dvipsnames,table,xcdraw]{xcolor}
\usepackage{array}
\usepackage{multirow}
\usepackage[normalem]{ulem}
\useunder{\uline}{\ul}{}
\usepackage{booktabs}
\usepackage{supertabular,booktabs}
\usepackage{longtable}
\usepackage{lscape}
\usepackage{cite}
\usepackage{url}
\usepackage{hyperref}
\usepackage{dirtytalk}
\usepackage{graphicx}
\usepackage{comment}
\usepackage{makecell}
\usepackage{svg}
\usepackage{amsmath}
\usepackage[caption=false]{subfig}
\graphicspath{{Images/}}
\usepackage[normalem]{ulem}
\useunder{\uline}{\ul}{}

\usepackage{tabularx}
\usepackage{balance}
\usepackage{nameref}
\newcommand{\numParticipants}{26} 

\begin{document}

\title{Post-pandemic Resilience of Hybrid Software Teams}

\author{
\IEEEauthorblockN{ Ronnie de Souza Santos}
\IEEEauthorblockA{ Cape Breton University\\
    Sydney, NS, Canada \\
    ronnie\_desouza@cbu.ca }
\and
\IEEEauthorblockN{ Gianisa Adisaputri }
\IEEEauthorblockA{ Dalhousie University\\
  Halifax, Canada \\
  gianisa@dal.ca }
\and
\IEEEauthorblockN{ Paul Ralph }
\IEEEauthorblockA{ Dalhousie University\\
  Halifax, Canada \\
  paulralph@dal.ca } 
}





\IEEEtitleabstractindextext{%
\begin{abstract}
\textit {Background}. The COVID-19 pandemic triggered a widespread transition to  hybrid work models (combinations of co-located and remote work) as software professionals' demanded more flexibility and improved work-life balance. However, hybrid work models reduce the spontaneous, informal face-to-face interactions that promote group maturation, cohesion, and resilience. Little is known about how software companies can successfully transition to a hybrid workforce or the factors that influence the resilience of hybrid software development teams.  
\textit {Goal}. The purpose of this study is to explore the relationship between hybrid work and team resilience in the context of software development. 
\textit {Method}. Constructivist Grounded Theory was used, based on interviews of \numParticipants{} software professionals. This sample included professionals of different genders, ethnicities, sexual orientations,  and levels of experience. Interviewees came from eight different companies, 22 different projects, and four different countries. Consistent with grounded theory methodology, data collection, and analysis were conducted iteratively, in waves, using theoretical sampling, constant comparison, and initial, focused, and theoretical coding.  
\textit {Results}. Software Team Resilience is the ability of a group of software professionals to continue working together effectively under adverse conditions. Resilience depends on the group's maturity. The configuration of a hybrid team (who works where and when) can promote or hinder group maturity depending on the level of intra-group interaction it supports. 
\textit {Conclusion}. This paper presents the first study on the resilience of hybrid software teams. Software teams need resilience to maintain their performance in the face of disruptions and crises. Software professionals strongly value hybrid work; therefore, team resilience is a key factor to be considered in the software industry.
\end{abstract}

\begin{IEEEkeywords}
team resilience, hybrid work, grounded theory, software team.
\end{IEEEkeywords}}

\maketitle

\IEEEdisplaynontitleabstractindextext

\IEEEpeerreviewmaketitle

\section{Introduction}\label{sec:introduction}

\IEEEPARstart{T}{he} mass transition to remote work amid the COVID-19 pandemic profoundly affected individuals, teams, and organizations. In the software industry, professionals abruptly shifted into ostensibly temporary home offices, which feel rather permanent two years later~\cite{brynjolfsson2020covid,craig2021dual,collins2021covid}. Often, COVID lockdowns erected physical and conceptual barriers between software team members, severing their emotional connections and triggering complex adaptations, rearrangements, and changes to team dynamics~\cite{ralph2020pandemic, ford2021tale}. 

The software industry has a history of geographically distributed teams. However, this was not enough to keep the pandemic from undermining the face-to-face interactions that are necessary to build cohesion and resilience, which are key antecedents of team performance~\cite{vcavrak2019resilience}. Now, as pandemic restrictions wax and wane, many software professionals decided to continue working remotely, motivated by more flexibility and improved work-life balance~\cite{chavez2020permanently,ralph2020pandemic,santos2022grounded}. Many organizations are, therefore, permanently transitioning to hybrid work models where individuals can choose to work in the office or other places depending on their needs ~\cite{barrero2021let, melin2021employees}.

Hybrid work is different from distributed software development because, in this work arrangement, which is becoming very popular after the pandemic, software team members can work together in an office on any given day while others work from home or other remote locations~\cite{santos2022grounded}. The high level of flexibility and interchangeability on how many people are at the office and when is the general characterization of hybrid work in the post-pandemic. Based on this dynamic, hybrid software teams are defined as teams working in fully flexible and erratic locations~\cite{smite2022future}, with individuals constantly changing from the office to working from home, depending on their needs.

Benefits of hybrid work include improved workplace accessibility for persons with disabilities; better work-life balance and flexibility for daily responsibilities (e.g., parenting); and reduced carbon footprint and costs from commuting and real estate~\cite{ralph2020pandemic, ford2021tale, chavez2020permanently}. Hybrid work also helps LGBTQA+ employees by giving them more control over identity disclosure~\cite{ford2019remote}. These benefits should improve equity and sustainability in the workspace. However, hybrid work still limits face-to-face interaction, which is critical for forging team resilience.

Software teams need resilience to remain effective in extraordinary circumstances~\cite{sharpwe}. Still, little is known about how software companies can successfully adopt a hybrid workforce strategy that provides professionals with the desired levels of flexibility in the post-pandemic. On the other hand, refusing to do so can drive the industry to a great resignation, with an increased number of software professionals quitting their jobs and seeking environments that are more innovative in terms of working models~\cite{barrero2021let, serenko2022great}. Therefore, in this study, we explore the characteristics of resilience in hybrid software teams aiming to support software companies in mitigating challenges resulting from the increasing number of professionals adhering to this work arrangement. 

The study of resilience in the context of hybrid software teams is pertinent because:

\begin{enumerate}
    \item the number of hybrid teams is rapidly increasing in the post-pandemic, and we cannot afford to have these teams disintegrate.
    
    \item remote work allowed equity-deserving groups to have access to opportunities and positions that were not available before (e.g., working remotely), and resilient hybrid teams are the key to maintain this achievement.
    
    \item we need to learn from the pandemic experience and leverage the recent crisis to expand our knowledge of maintaining team performance during disruptions, ensuring that software teams can continue working during disasters, emergencies, and ongoing crises.
    
    \item while much research examines software teams' productivity and well-being in the pandemic era~\cite[e.g.]{ralph2020pandemic}, little research has included resilience as an outcome variable.
\end{enumerate}

However, `hybrid' refers to a wide spectrum of work arrangements~\cite{smite2022future} from `you can work from home once in a while if needed' to `half the team is fully remote, and the other half comes in one day a week, but not the same day'. No previous studies investigated how this spectrum affects team resilience (or other outcome variables commonly studied in software engineering). We, therefore, embarked on a grounded theory study to investigate the following research question.

\smallskip
{\narrower \noindent \textit{\textbf{Research Question:} What is the relationship, if any, between hybrid work models and software team resilience?}\par}
\smallskip

From this introduction, this study is organized as follows. In Section 2, we discuss existing studies about resilience and related works. In Section 3, we describe the method conducted in this study. In Section 4, we present our findings which are discussed in Section 5, along with the implications and limitations of this study. Finally, Section 6 summarizes the contributions of this study and opportunities for future research.

\section{Background} \label{sec:RelatedWork}
Studies on psychological resilience have been gaining interest since 1970, with an early focus on the positive adaptations of children, youths, and families facing significant and prolonged adversities and how they manage to overcome the adversities and reach competent functioning in various aspects of their lives as adults~\cite{masten2018resilience, bonanno2015temporal}. As a psychological trait, resilience is considered an individual’s ability to persevere and thrive through challenges~\cite{block1996iq}. However, current research on resilience emphasizes the interactions of multiple adaptive systems (i.e., biological, social, psychological, and ecological) that influence individuals’ abilities to cope, respond and thrive to challenges~\cite{masten2021multisystem, ungar2020resilience}.

With the current global challenges and prominent team-based structures in organizations, workplace issues became prominent adversities that affect not only individuals but also team performance, efficacy, and well-being~\cite{flint2017team}. These challenges range from intense work pressures~\cite{badu2020workplace}, technical difficulties, feeling of isolation, trouble building trust~\cite{kirkman2002five, wagner2007filling} to differences in values between members, and cultural and language barriers~\cite{holmes2005small}. For instance, nurses experience mental health issues due to workplace stress; however, this effect is moderated by positive work interactions and organizational support\cite{badu2020workplace}. Similarly, positive emotions help teams cope with workplace pressures and increase their performance~\cite{meneghel2016feeling}. However, despite the apparent link between individual capability and team performance, a team's capacity to overcome adversity is independent of individual members’ resilience~\cite{flint2017team}. 

Team resilience has been defined as a psychosocial process, a capacity, an emergent state, or an outcome when a team faces significant adversity~\cite{chapman2020team}. For our purposes, \textit{team resilience} is ''a team's capacity to withstand or recover from adverse events (i.e., events that may lead to losses or breakdown of independent team processes)'', which we conceptualize as an emergent team state that results from preparative, adaptive, and reflective team processes and which is demonstrated by a persistence, recovery, or growth trajectory of team functioning following exposure to adversity.~\cite[p.186]{hartwig2020workplace}    

Team resilience is influenced by numerous factors, including collaborative culture, shared empathy~\cite{teekens2021shaping}, interpersonal trust, group confidence and efficacy~\cite{pavez2021project}, and positive leadership styles~\cite{vera2017may}. Highly resilient teams exhibit more positive emotions, team performance~\cite{meneghel2016feeling}, mastery~\cite{morgan2013defining}, and individual resilience~\cite{bennett2010team}. Team resilience also increases individual motivations and work commitment, which in turn strengthens organizational resilience~\cite{teekens2021shaping}.

In software engineering, resilience is commonly associated with the capacity of a component (i.e., system, server, network, database) to quickly recover from disruptions and continue to operate~\cite{murray2017resilience}. In this sense, while there are a significant number of studies available on systems' resilience~\cite{patriarca2018resilience}, the study of resilience from a human perspective is not common, although software professionals and software teams are frequently required to adapt to cope with changes in the software development processes ~\cite{li2015makes, santana2015software}. 

Considering human aspects of software engineering, resilience has been mentioned in some studies~(e.g.~\cite{marinho2018managing,deshpande2016remote,cruz2015forty}); however, it has never been the main topic of investigation. Hence, discussions about individual and team resilience in software engineering are scarce. Some studies emphasized the need to recognize the uncertainty of software projects and prepare to adapt to unexpected situations~\cite{marinho2018managing,deshpande2016remote}. In addition, individual resilience is assumed to relate to job satisfaction; however, no further investigations were conducted~\cite{cruz2015forty}.

Since COVID-19, more studies have discussed how software teams and software professionals were adapting to the crisis~\cite{nazir2022adapting, griffin2021implementing}. As many companies developed virtual group activities (e.g., games, chats) to increase social interaction among professionals working remotely ~\cite{desouzasantoschase}, a supportive working environment along with effective communication are reported to be the key to improving professionals’ resilience~\cite{marinho2021happier, jay2021has}. However, similar to what happened in studies developed before the pandemic, such findings require further investigations. 

\section{Method} \label{sec:method}
We applied Constructivist Grounded Theory (CGT) based on semi-structured interviews following commonly-used guidelines~\cite{charmaz2014constructing,stol2016grounded,ralph2020empirical}. Grounded Theory refers to a family of research methods widely applied in social sciences to generate theories inductively from observations~\cite{glaser1978theoretical}. Its main characteristic is the use of detailed rounds of data collection and analysis---including inductive coding, memoing, constant comparison, and theoretical sampling---allowing concepts to emerge rather than fitting into pre-conceived theories, concepts, or categories~\cite{glaser1978theoretical}. Grounded Theory is particularly suitable for investigating software engineering's social, cultural, and human aspects~\cite{stol2016grounded}. CGT is similar to ``Classic'' Grounded Theory but modified to fit a constructivist epistemology; it focuses on investigating real-life environments to identify concepts that explain behaviors and experiences~\cite{charmaz2014constructing}.  

\subsection{Participants} \label{sec:participants}

CGT provides a deep, nuanced understanding of a real-life phenomenon based on the experience of a small group of individuals~\cite{charmaz2014constructing}. For this study, we focused on professionals who work in the kind of hybrid environment where everyone on their team lives in the same city or region where their physical office is, but are free to go to the office or work from home as they like. We focused on this situation because it clearly differs from global, distributed, or remote-first software engineering. We began by inviting professionals that participated in a previous study on a similar topic \cite{santos2022grounded} (i.e., convenience sampling~\cite{baltes2022sampling}). We then asked participants to suggest other professionals who might be interested in the study (i.e., snowball sampling). We interviewed this second group of participants beginning in the third round of interviews. 

Another kind of sampling that is important in CGT is theoretical sampling~\cite{charmaz2014constructing}. For this study, theoretical sampling was more about what questions we asked than who we interviewed. As our findings and theory took shape, we modified our interview guide to collect data that filled in the gaps. 

Previous studies on how software professionals experienced working from home during the pandemic demonstrated that some groups of individuals were facing more challenges than others (e.g., women and people with disabilities)~\cite{ralph2020pandemic}. Therefore, we aimed to explore as many experiences as possible. We increased the variety of our sample with participants from different genders, backgrounds (e.g., ethnicity, sexual orientation), skills (e.g., specialties, roles), work experience, team structures (e.g., number of co-workers), and project domains (e.g., web, mobile, desktop). We also considered the location of participants; however, the convenience sampling influenced on the variety of countries in the sample. In this sense, we used the location of clients, if applicable, to maximize the diversity of locales in the sample since many cultural aspects related to the clients directly influence the routine of software teams, for instance, defining strategies, methods, and metrics. Maximizing sample diversity like this is called \textit{heterogeneity sampling}~\cite{baltes2022sampling}.

\subsection{Data Collection}

We conducted semi-structured interviews to investigate participants' experiences developing software in hybrid contexts; for instance, working with high levels of flexibility and interchangeability~\cite{santos2022grounded, smite2022future}. Interviews were conducted in six rounds. We began with four participants per round and increased the number of participants in the last round as answers started to become repetitive (saturating). The rounds of data collection were systematically improved and adapted, guided by our theoretical sampling, including changing interviewing styles (e.g., asking pre-defined questions or simply listening to participants describing their experiences), participant profiles, and level of detail (e.g., strategically asking key participants about unsaturated aspects of the emerging theory).

Interviews were conducted online following a pre-established interview guide and using the participant’s choice of virtual meeting service (mostly MS Teams and Google Meet). We collected data until reaching saturation, which happened when new interviews were not adding novelty to the obtained findings ~\cite{aldiabat2018data}. We finished with a total of \numParticipants{} interviews ranging from 22 to 73 minutes. This produced 13 hours and 38 minutes of audio and 241 pages of transcripts. Participants were asked about their experiences working in hybrid environments, the advantages and challenges of hybrid work, and the impacts of hybrid work on their teams. 

\subsection{Data Analysis}
We began data analysis by transcribing and line-by-line open coding the first four interviews to identify emerging concepts and their properties~\cite{charmaz2014constructing}. We repeated this strategy for each round by applying open coding and comparing the new emerging codes to our collection of codes in a process of constant comparison. Audio recordings helped give meaning to the codes, while memos and notes that were produced during the interviews helped in the identification of key points and pertinent quotes. 

Around the thirteenth interview, we began categorizing codes and establishing preliminary connections among them using focused coding; that is, by making repeated analyses and comparisons among emerging categories~\cite{charmaz2014constructing}. This process continued during the fourth round of data collection, and when these interviews were integrated into the analysis the core category emerged from our interconnections among categories. 

Following this, we conducted theoretical coding, a process focused on iteratively rearranging the categories around the core category until they stabilized and the connections among them were established~\cite{charmaz2014constructing}. At this point, the final round of interviews confirmed that our conclusions from the participants' perspectives were clearly explained. Simultaneously, we conducted one round of member checking by soliciting feedback on our findings from willing participants. The member checking resulted in a) positive comments on the findings; b) shared perception that the main concepts were included; c) small adjustment in the codes naming; d) confirmation that the theory was saturated, e.g., there was nothing new to be incorporated into the theory. 

\subsection{Ethics}
We generally followed the norms of the ethics committee at the third author’s university, which approved this research. Each professional invited to participate in this research agreed to provide data and to be interviewed and recorded. Before each interview, the interviewer explained the goal of the research and its relevance to the software industry. Participants were asked to verbally agree with the audio recording and were guaranteed data confidentiality and de-identification of their quotes. No participants withdrew from the research.

\section{Findings} \label{sec:results}
In this section, we first report participant demographics and then present the categories that emerged from the analysis and explain how they are interconnected. Some of the quotations used below have been translated into English by the authors. Some quotations may read awkwardly as we have endeavored to transcribe and translate them as accurately as possible. 

\subsection{Demographics}
As described in Section~\ref{sec:participants}, software professionals were selected to participate in this study using a theoretical sampling approach~\cite{glaser1978theoretical}. During the first round, we purposively interviewed professionals that participated in a previous study on remote work to understand how their experience evolved over time. In the following rounds, we interviewed participants focusing on maximizing sample diversity to support developing a theory that unites different perspectives, experiences, and realities.

Through this approach, our sample was composed of \numParticipants{} software professionals who are consistently working on hybrid teams; that is, the current norm in their team is that professionals can choose to work at the office or remotely with no restrictions. In this strategy, to keep the sample aligned with our primary goal, we interviewed only professionals that explicitly affirmed that either they or someone in the team developed the habit of regularly going to the office; thus, allowing us to explore hybrid work. Table 1 summarizes the sample of participants in this study.

The sample included individuals from many different roles, including  programmers, software QAs/testers, designers, functional analysts (e.g., requirements analysts), and software project managers. The variety of backgrounds allowed us to understand how hybrid work affects different software development activities and their outcomes. Furthermore, we interviewed professionals working on different system domains (e.g., web, mobile, and desktop), along with different levels of experience (e.g., from Junior professionals with one year of experience to experienced professionals working in software development for over ten years). 

Considering practices, processes, culture, and organizational contexts, we interviewed professionals from eight different companies and 22 different projects, which allowed us to explore hybrid software development from a variety of scenarios. This included projects using Scrum, Kanban, and ad-hoc methods. Moreover, the companies were located in four different countries, and the interviewees were working for clients from seven countries. The company's country and the client's country are relevant information in this study because organizational and cultural aspects related to the location might affect the way that hybrid work is experienced by the software team.

In addition, by recognizing the lack of diversity in software engineering research~\cite{rodriguez2021perceived,albusays2021diversity}, we included individuals from equity-deserving groups. We believe that understanding their experience is crucial for improving software processes, especially when remote work affects these professionals differently~\cite{ralph2020pandemic,ford2019remote}. In this sense, our final sample included the following:
\begin{itemize}
\item 35\% women, including cisgender and transgender women.
\item 4\% individuals with disabilities (one person with impaired vision).
\item 42\% non-white individuals, including Black and Indigenous professionals.  
\item 23\% non-straight individuals, including gay, lesbian, pansexual, and asexual professionals.
\end{itemize}

\begin{table}
\centering
\caption{Demographics}
\renewcommand{\arraystretch}{1}
\label{tab:Demographics}
\begin{tabular}{llr}
\toprule
\multirow{2}{*}{ \textbf{Gender} } & Men & 17$^a$\\
& Women & 9$^b$ \\ \midrule 
\multirow{2}{*}{ \textbf{Disability} } & Without & 25\\
& With & 1 \\ \midrule 
\multirow{3}{*}{ \textbf{Ethnicity} } & White & 15 \\
& Black & 9 \\
& Indigenous & 2 \\ \midrule 
\multirow{5}{*}{ \textbf{Sexual Orientation} } & Straight & 20 \\
& Gay & 2 \\
& Lesbian & 1 \\
& Pansexual & 2 \\
& Asexual & 1 \\ \midrule 
\multirow{5}{*}{ \textbf{Role} } & QAs/Testers & 10 \\
& Programmers & 8 \\
& Designers & 4 \\
& Managers & 2 \\
& Functional Analysts & 2 \\ \midrule 
\multirow{4}{*}{ \textbf{Experience} } & Senior & 8 \\
& Mid-level & 7 \\
& Junior & 7 \\
& Principal & 4 \\ \midrule 
\multirow{4}{*}{ \textbf{Domain} } & Web & 12 \\
& Multiple & 5 \\
& Mobile & 5 \\
& Desktop & 4 \\ \midrule 
\multirow{4}{*}{ \textbf{Location} } & Brazil & 21 \\
& US & 3 \\
& Portugal & 1 \\
& Japan & 1 \\ \midrule 
\multirow{5}{*}{ \textbf{Client Location} } & US & 13 \\
& Brazil & 8 \\
& Canada & 1 \\
& Japan & 1 \\
& Portugal & 1 \\
& South Korea & 1 \\
& UK & 1 \\
\bottomrule
\end{tabular}

\flushleft
\footnotesize{Notes: $^a$16 cisgender, 1 transgender; $^b$7 cisgender, 2 transgender}
\end{table}


\subsection{Overview of Themes}

The core category that emerged from the analysis was \textit{Hybrid Software Team Resilience}, which is defined as team resilience (as described above) in the context of software development. Hybrid Software Team Resilience is associated with five themes---Belonging, Proactiveness, Conflict management, Organizational support, and Diversity---which are defined in Table~\ref{tab:themes}. Each theme is associated with two or more subcategories, summarized by Table~\ref{tab:subcategories}. Next, we discuss each theme in more detail.

\begin{table*}[t!]
\caption{Theme definitions and illustrative quotations*}
\label{tab:themes}
\renewcommand{\arraystretch}{1.3}
\begin{tabularx}{\textwidth}{p{1.8cm}XX}
\toprule
\textbf{Construct} & \textbf{Definition} & \textbf{Example Evidence} \\
\midrule
Software team resilience & The ability of a group of software professionals to continue working together effectively under adverse conditions. & ``[in] a resilient hybrid team \dots even in the absence of one or more members, activities will continue''~(P15) \\
Belonging & The feeling of being part of the team and included in the team structure. & ``if you feel part of the team, you do your best working with those people''~(P13) \\
Diversity & Variety in team members' personal, professional, and technical backgrounds. & ``they are all white men \dots so they struggle a little in relation to the diversity''~(P13) \\
Conflict management & How well the team resolves or avoids disagreements & ``but unfortunately they fight a lot; a matter of ego''~(P13) \\
Organizational support & Resources that provide employees with a similar work experience regardless of location. & ``We have a laboratory at the office but every QA and some developers have the hardware at home.''~(P11) \\
Proactiveness & The degree to which team members initiate---and take responsibility for---important tasks without prompting & ``people started realizing that we are a team, so we need to deliver everything together.''~(P12) \\
\bottomrule
\end{tabularx}
\flushleft
*more supporting quotations are available (see \textit{\nameref{sec:DataAvailability}})
\end{table*}

\begin{table*}[t!]
\caption{Subcategory definitions and illustrative quotations*}
\label{tab:subcategories}
\renewcommand{\arraystretch}{1.3}
\begin{tabularx}{\textwidth}{l>{\raggedright\arraybackslash}p{2.1cm}XX}
\toprule
\textbf{Theme} & \textbf{Subcategory} & \textbf{Definition} & \textbf{Illustrative Quotation} \\
\midrule
\multirow{3}{*}{ Belonging } & Consistent interaction & The frequently and consistently interacting with teammates & ``we maintain an ongoing call all day long as if we were at the office''~(P04) \\
& Internal feedback & The frequency and effectiveness of feedback from teammates, task results, or managers (e.g. in a retrospective meeting) & ``He gave me good feedback (...) Then told me what should I keep focusing.''~(P26) \\
& Mutual support & The assistance that individuals receive from teammates in difficult moments. & ``When I told them that I had a bleeding problem ... they said, don't worry, we'll take care of your tasks. Go rest.''~(P19) \\
\midrule
\multirow{5}{*}{ Proactiveness } & Update seeking & The tendency to maintain awareness of the status of one's project and team members & ``I start my day by checking the messages from the day before''~(P02) \\
& Autonomy & The freedom to take action and make decisions that can affect the whole team. & ``each one decides what is the best moment [to take action]''~(P12) \\
& Shared responsibilities & The attitude of collective ownership in the project, its artifacts, and its outcomes. & ``tasks are not associated to a person, but to the product''~(P12) \\
& Task interchangeability & The degree to which any team member can complete any of the team's tasks. & ``the company has a culture of swapping activities to avoid knowledge islands''~(P07) \\
& Quality focus & The attitude that all team members are equally responsible for the quality of the product. & ``the team understands that the responsibility for quality is not only the QA’s.''~(P12) \\
\midrule
\multirow{4}{*}{ \parbox{1.5cm}{Conflict management} } & Project awareness & Team members' knowledge of the project and its status & ``you need a general level of knowledge about the project and the deliverables and the development and everything else''.~(P01) \\
& Defined routine & The extent to which the team shares a common schedule (e.g. stand-up meetings at 9 am). & ``we’ve got to the point that we built this great routine''~(P05) \\
& Atomic tasks & The extent to which the tasks are well-defined and sufficiently independent that team members can work in parallel without conflict. & ``I miss having a greater division of modules and an explanation among these modules.''~(P14) \\
& Leadership transparency & The team is well-informed about the decisions and actions of management. & ``the person working remotely suffered a lot because they [the leaders] didn't document decisions and that person ended up confused.'' ~(P08) \\
\midrule
\multirow{3}{*}{ \parbox{1.5cm}{Organizational support} } & Location flexibility & The degree to which employees can choose \textit{where} they work (i.e. from home or at the office). & ``there are people who stated clearly and openly that if the company force them to return to in-person, they would resign''.~(P05) \\
& Resource convenience & The services and amenities that allow professionals to have access to project-related technologies (e.g., delivery of equipment). & ``a delivery boy picks up the phone at [COLLEAGUE NAME]’s house and delivers it to me”~(P17) \\
& Commuting assistance & The strategies that can ease the access of professionals to the office
when necessary (e.g., rides, passes or transport benefits) & ``so, there is a specific area focused only on tickets [purchase], for flights.''~(P05) \\
\midrule
\multirow{2}{*}{ Diversity } & Diverse Needs Accommodation & The strategies used by the team to include and support the unique needs and circumstances of its members (e.g. not scheduling meetings when children need to be picked up from school) & ``The routine has changed a little because now we have an organized routine and I only have one car for me, my wife, my boys and so on.''~(P09) \\
& LGBTQIA+ visibility & The actions taken to assure that LGBTQIA+ professionals will continue comfortable in working with the team despite the work environment. & ``I get anxious that people learn that I'm trans. It makes me afraid of experiencing any type of violence, you know?'' (P03) \\
\bottomrule
\end{tabularx}
\flushleft
*more supporting quotations are available
\end{table*}

\subsection{Resilience}
As explained in Section~\ref{sec:RelatedWork}, resilience is ``a team's capacity to withstand or recover from adverse events''~\cite[p.186]{hartwig2020workplace}. In other words, software teams need resilience to maintain their performance in the face of adverse situations. Considering hybrid work, these adversities  are related to disruptions in several aspects that affect software development activities, such as communication issues~\cite{miller2021your}, coordination difficulties~\cite{santos2022grounded}, decrease of involvement and cohesion~\cite{miller2021your, desouzasantoschase}, and problems of environment adaptation~\cite{bezerra2020human}.

Software professionals have a positive attitude towards hybrid work. Participants reported that the flexibility resulting from hybrid work increases their satisfaction. However, a hybrid environment requires their team to be more resilient than they needed to be before the pandemic (in-person) or even during the pandemic (completely remote) because the work structure in a hybrid environment is too dynamic as it can assume a wide range of configurations that can frequently change. There is no longer everybody at the office or everybody remote. 

The constant changes in the environment require the teams to be resilient and rapidly adapt to the situations at hand before disruptions affect communication schemes, coordination structures, and cohesion levels. In this sense, our participants pointed out five main factors that together support hybrid team resilience, they are: belonging, diversity, conflict management, organizational support, and proactiveness.


\subsection{Antecedents of Resilience in Hybrid Software Teams}
 The following five factors are the pillars that support teams to overcome these challenges. Figure 1 illustrates how the resilience of hybrid teams is affected by these factors. Table 2 shows the chain of evidence obtained from our participants.

\subsubsection{Belonging}
Like remote work, hybrid work creates physical barriers between team members, which reduces interactions among professionals. Software development is highly dependent on such interactions (e.g., discussing problems, ideas, and solutions). Many participants reported that a sense of belonging is an important aspect of team resilience as it gives them the feeling that they are part of a team and not simply working with a group of individuals. Therefore, a sense of belonging supports team resilience as individuals commit to the team and work together towards overcoming challenges.

The participants discussed three factors related to a sense of belonging:
\begin{itemize}
    \item \textit{Consistent Interaction}: Sense of belonging depends on team members interacting often to build team cohesion.
    \item  \textit{Internal Feedback}: Sense of belonging depends on the frequency and effectiveness of the feedback received from teammates and leaders, which helps professionals to feel included and valued. 
    \item \textit{Mutual Support}: Sense of belonging depends on how professionals are assisted by their teammates, including assistance on matters that are not related to the work.
\end{itemize}

\subsubsection{Proactiveness}
Some interviewees expressed that hybrid work increases the risk of loss of connection, power outages, and other issues that can keep individuals from working. In this work context, they need to be proactive and act in anticipation before problems arise or intensify. 

The interviewees pointed out five factors that contribute to team proactiveness:
\begin{itemize}
    \item \textit{Update Seeking}: Proactiveness depends on strategies to maintain shared mental models as individuals develop the habit of trying to be constantly updated. 
    \item \textit{Autonomy}: Proactiveness depends on team members being empowered to make responsible and informed decisions that affect co-workers or the whole team. 
    \item \textit{Shared Responsibility}: Proactiveness depends on professionals taking responsibility for activities that are not necessarily assigned to them, for instance, when a co-worker is facing difficulties. 
    \item  \textit{Task Interchangeability}: Proactiveness depends on tasks being easily interchangeable, i.e., not depending on specific individuals; they are easily swapped when necessary. 
    \item \textit{Quality Focus}: Proactiveness depends on every team member assuming the responsibility for the quality of the product, i.e., everyone is committed to the quality, not just testing professionals.   
\end{itemize}

\subsubsection{Conflict Management}
Hybrid work is highly dependent on virtual environments. The work in these environments is susceptible to many conflicts and disagreements that trigger disruptions in the team. Participants described that effective conflict management is the key to recovering from several disruptions in the hybrid team. 

By managing conflicts, the team would avoid spending energy solving disruptions caused by misunderstandings or miscommunications and focus on overcoming other types of challenges and adversities. Four factors can be managed to decrease conflicts in hybrid contexts and increase team resilience:

\begin{itemize}
    \item\textit{Project Awareness}: Conflict management is supported by providing individuals with a common body of knowledge about the project and its status to avoid misunderstanding and misinterpretation.
    \item \textit{Defined Routine}: Conflict management is supported by creating basic routines and common schedules for the team (e.g., synchronous meetings, core work hours) to avoid process conflicts (e.g., coordination).
    \item  \textit{Atomic Tasks}: Conflict management is supported by planning and organizing tasks in a way that the flow of execution is well-defined and independent to avoid task conflicts. 
    \item \textit{Leadership Transparency}: Conflict management is supported by communicating decisions to professionals to avoid misinformation and relationship conflicts.       
\end{itemize}

\subsubsection{Organizational Support}
One of the main challenges of hybrid work reported by many participants is how the software development environment can frequently change, which increases the complexity of maintaining a structure to have a team effectively working in the office and remotely at the same time. Therefore, team resilience strongly depends on the resources and support provided by the organization. Software companies are required to develop a culture that embraces hybrid work and provides employees with similar work conditions despite where they are.

According to the participants, organizational support not only helps teams to be more resilient but also influences turnover decisions, for instance, when professionals have to decide whether to stay in the company or accept another job. Three elements can enhance organizational support:

\begin{itemize}
    \item\textit{Location Flexibility}: Professionals expect organizations to allow them to freely decide where they will work while also providing them with resources to perform their tasks in person or remotely. 
    \item \textit{Resource Convenience}: Professionals expect organizations to develop strategies that provide them remote access to specific resources required by the project, such as exclusive hardware, licenses, or credentials.
    \item \textit{Commuting Assistance}: Professionals expect organizations to assist them to commute to the office when necessary. 
\end{itemize}

\subsubsection{Diversity}
Dealing with the challenges of hybrid work can be complex, and sometimes greater cohesion and support are insufficient to overcome problems and recover from them. In this context, team diversity is another factor that can help teams to cope with difficulties. Participants discussed that diverse teams are more resilient because diversity increases the chances of having someone with the expertise necessary for the adaptation to be successful. 

Diverse software teams are composed of a mixture of professionals not only from different technical expertise and experience levels; they also embrace a variety of backgrounds, ethnicities, genders, cultures, ages, and sexual orientations, which provides the team with a wider range of life experiences and perspectives to solve problems and implement software solutions. Homogeneous teams are less resilient because individuals tend to process the challenge in a similar way, due to their similar backgrounds; therefore, being less creative in searching for solutions to overcome adversities. 

Our participants identified two factors that can increase diversity in hybrid software teams:

\begin{itemize}
    \item\textit{Diverse Needs Accommodation}: Diversity depends on strategies to assist experienced professionals who are parents and might need additional support to cope with the hybrid model. 
    \item \textit{LGBTQIA+ Visibility}: Diversity depends on creating a safe and inclusive environment for LGBTQIA+ individuals, in particular, giving them more control over identity disclosure.
\end{itemize}

\section{Discussion} \label{sec:discussion}

\subsection{Reintegrating the Emerging Theory Into Existing Literature}

While our initial model touches on many well-researched phenomena, in this section, we focus on reconciling it with a handful of influential theories.

\subsubsection{Group Development} \label{sec:groupDevelopment}

Groups develop in at least four phases~\cite{tuckman1965developmental}:
\begin{enumerate}
    \item \textit{Forming:} Individuals get to know each other, work mostly independently, and avoid conflict.
    \item \textit{Storming:} As the individuals become comfortable and work \textit{together}, more intragroup conflict arises. The individuals become a group in the sense that they work \textit{together}, not just on the same project.
    \item \textit{Norming:} The group resolves its internal disagreements by establishing norms; and matures into a \textit{team}; that is, a group that works effectively because it shares a common identity, goals, and mental models~\cite{gren2017group}. 
    \item \textit{Performing:} Having resolved their disagreements and become cohesive, the team exhibits high productivity and strong performance. Disruptions (e.g., crises, or adding new team members) can send the team back to the storming stage.
\end{enumerate}

A \textit{mature} group is a group that has reached the performing stage. For practical purposes,  most organizational leaders and managers want mature teams. Forging a collection of individuals into a mature team is obviously important because an immature team is, by definition, ineffective. Mature is good---full stop. 

The \textit{belonging}, \textit{proactiveness} and \textit{conflict management} themes described above are all related to the concept of group maturity. 
Members of a mature team feel a sense of belonging and connection to their team~\cite{carron2007team}, which is sometimes called \textit{team cohesion}. Cohesive teams work together regularly, share goals, feel like they are in it together, and resolve conflicts constructively~\cite{wheelan1996validation}---that is, mature teams have good conflict management. Mature teams are also more proactive~\cite{sjovold2007systematizing}. 

However, the themes that emerged from this study do not cover the concept of maturity completely. Mature groups also have good communication and coordination, effective decision-making processes, and flexible procedures~\cite{schermerhorn2011organizational}. Some research suggests that the move to remote work amid COVID restrictions specifically undermined coordination, communication, and trust~\cite{santos2022grounded}.

Many aspects of group maturity are also associated with resilience. For example, team cohesion (i.e., belonging) promotes team resilience~\cite{alliger2015team,mathieu2015modeling}. Proactiveness, including collaborative culture and shared understanding of others' responsibilities, increases resilience by creating similar goals and enhancing work motivations to achieve said goals~\cite{teekens2021shaping}. Furthermore, teams with better collectivist conflict management are more likely to have better engagement, and thus will perform better in challenging situations~\cite{stoverink2020bouncing}. Other traits, such as transparent decision-making, flexibility in procedure modification, efficacy, group autonomy and awareness of team members' tasks and challenges also enhance team resilience in adversities~\cite{stoverink2020bouncing,mcewen2018measure, cook2020team}.              

Indeed, group maturity, cohesion, and resilience emerge together from the storming and norming phases, which are driven by regular, often informal, often spontaneous face-to-face interactions~\cite{wheelan2020creating}. Norming happens because people work \emph{together}, not individually. A hybrid work model may reduce precisely the kind of interactions that turn individuals into mature, cohesive, resilient teams. 

\subsubsection{Diversity} \label{sec:diversity}
The relationship between workplace diversity and team dynamics cannot be reduced to `diversity is good' or `diversity is bad'~\cite{harrison2007s}. Diversity \textit{in experience and expertise} increases team creativity and problem-solving abilities~\cite{cummings2004work,harrison2007s}. However, the effect of highly visible demographic diversity (e.g., race, age, gender, disability status) on team dynamics is more complicated~\cite{pelled1996demographic,harrison2007s,fassinger2008workplace}. 

High-visibility demographic diversity directly affects interactions at work, in both formal and informal spaces, by triggering the cognitive process of categorizing us versus others~\cite{wentling1998current}. People from diverse populations are often invisible and excluded from the workplace and social networks~\cite{fassinger2008workplace}. This separation, combined with preferences to interact with people from the same subgroups, may reduce team cohesion~\cite{harrison2007s}. 
 
Under some conditions, however, highly visible diversity can have a neutral or positive effect on the team dynamics~\cite{wentling1998current,harrison2007s, harvey2013different}. When team members’ views, goals, and attitudes are aligned, a variety of perspectives within a group may improve creativity~\cite{harrison2007s}. When a team is already cohesive, with high inclusivity, appreciation for other team members, frequent communication, and an effective task process, more diversity may reduce conflict~\cite{harvey2013different,wentling1998current}. When team members feel like they belong, are respected, and are listened to, diversity may increase  coordination and therefore performance~\cite{harrison2007s}. 


\begin{figure*}[t]
\centering
\centerline{\includegraphics[width=0.75\textwidth]{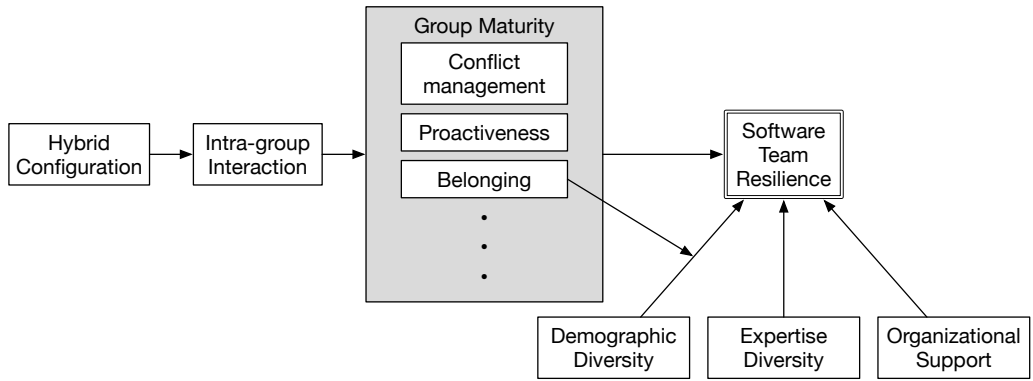}}
\caption{Theory of resilience in hybrid software engineering}
\label{fig:theory}
\end{figure*}

\subsection{Towards a Theory of Hybrid Software Team Resilience}

Armed with the insights from prior research described above, we can organize our themes into a theory of software team resilience shown in Figure~\ref{fig:theory}. 

Starting on the left, a team's hybrid work model can be configured in many different ways. Our participants suggested that the two main parameters of this configuration are: (1) the number of days and hours that software professionals are required to be in the office; and (2) the number of team members in the office at the same time. Different combinations result in different opportunities for the kind of spontaneous, informal interactions that drive group maturation. For example: 

\begin{enumerate}
    \item Most of the group works remotely most of the time. There are rarely two people in the office simultaneously. Group members work independently and check in only when forced to do so by a problem they cannot solve on their own. This configuration leaves little opportunity for spontaneous, informal interactions.
    \item All group members observe \textit{core work hours}, 9--5, Tuesday--Thursday and can work their remaining hours whenever they wish. This configuration provides many opportunities for spontaneous, informal interactions.
    \item Group members can work \textit{wherever} they wish, but all keep a joint video call open from 9 am to noon, Monday-Friday. This configuration provides many opportunities for spontaneous, informal interactions.
    \item Some group members work together in the office, while others work remotely except for mandatory meetings. This configuration provides many interaction opportunities for some members while excluding others.
\end{enumerate}

These examples illustrate how the configuration of the hybrid work model affects the degree of \textit{distance}: ``a difference in position or level between entities that requires effort to traverse to accomplish a software development task''~\cite[p. 210]{bjarnason2016theory}. Some configurations minimize distance, while others exacerbate distance. We posit that distance-minimizing configurations should better support group maturation---hence the causal arrow from \textit{hybrid configuration} to \textit{intra-group interaction} in Figure~\ref{fig:theory}. 

However, the minimum distance is achieved by inflexible, fully co-located work models. Practically speaking, organizations will want to optimize the hybrid configuration for multiple objectives, including cost, flexibility and distance, which is why we use \textit{hybrid configuration} rather than \textit{distance} as our leftmost construct.

Next, as explained in Section~\ref{sec:groupDevelopment}, intra-group interaction is an important antecedent of group maturity because the storming and norming phases of group development are driven by often spontaneous, informal interactions. 

We can think of Group Maturity as a kind of omnibus latent variable that includes much of what makes a software team good (other than technical skills). As explained in Section~\ref{sec:groupDevelopment}, mature teams are proactive and have good conflict management and cohesion (that is, each team member feels a sense of belonging to the team). Figure 1 shows conflict management, proactiveness, and belonging inside group maturity to indicate that these are interrelated dimensions of maturity.\footnote{We considered using UML symbols for composition or aggregation but felt that they would imply properties related to object-oriented programming that does not necessarily make sense here.}

The three dots inside the group maturity box indicate that group maturity includes several other dimensions (e.g., coordination, communication, and decision-making effectiveness) that have been established by prior research but were not highlighted by our participants. Practically speaking, an instrument to measure group maturity, such as the Group Development Questionnaire~\cite{wheelan1996validation}, should cover all of the major dimensions of maturity. Indeed, Wheelan and Hochberger~\cite{wheelan1996validation} list 40 characteristics of mature groups. 

Many of these dimensions of maturity are likely correlated and trying to tease out causal direction within the maturity construct (e.g., do feelings of belonging cause good decision making or does making good decisions make people feel like they belong, or are they both caused by good communication?) may not be particularly fruitful. From a software engineering perspective, what matters is that group maturity is the key driver of software team resilience. 

Other drivers of software team resilience include organizational support and the diversity of the team. However, as we saw in Section~\ref{sec:diversity}, the relationship between diversity and resilience is complicated. Diversity of expertise should increase resilience since the team will know how to handle more challenges. However, the relationship between demographic diversity and team resilience is moderated by belonging. In mature, inclusive teams where members all feel like they belong to the team and the team appreciates them, more demographic diversity should improve the team's resilience. However, in less mature groups with us--them dynamics, where some members feel excluded, higher demographic diversity may reduce team resilience. 




\subsection{Implications}
These research can support practitioners in dealing with problems arising from hybrid work in the software industry. We demonstrated that resilience is strongly related to team maturity. Software teams needs to improve their resilience to maintain productivity and performance in face of adversities. In this sense, we propose the following recommendations: 

\begin{itemize}
    \item\textit{Identify the hybrid configuration that better suits the team}: There are several possible configurations of hybrid models for a software team, depending on the number of people working remotely and the amount of time the team spends together (synchronously). Therefore, practitioners need to experiment and define a suitable hybrid configuration for their teams focusing on spontaneous and informal interactions; however, respecting professionals' restrictions and needs. We recommend that practitioners consider the characterization of a good hybrid configuration reported by the majority of participants of this study, which should include flexible days at the office chosen by professionals instead of mandatory days (e.g., every Wednesday), and the number of days should be based on commuting support provided by the organizations.
    \item\textit{Define a common routine for the team}: Independently of the hybrid configuration, software teams need a common routine that allows the team members to connect and discuss the work activities. This can be achieved by well-established practices in the industry, such as daily meetings and retrospectives, or by the definition of new practices. We recommend some practices identified in this study: joint video call at a specific time of the day, group call while working, specific moments for status sharing during the day, and checking backlog status as the first task of the day. 
    \item \textit{Improve team diversity}: Diverse teams are more creative; therefore, more prepared to face problems and recover from crises. For many years the software industry has been recognized for its lack of diversity. Now, hybrid teams are demonstrated to be more effective when professionals from different backgrounds are working together. We recommend that decisions on team composition are made not only on the technical background (e.g., level of experience, expertise, or education), but also on demographic diversity, including gender, race, culture, and sexual orientation. 
    \item \textit{Create strategies to increase inclusiveness}: Professionals are more committed when they feel included by their team members. The sense of belonging is essential for the resilience of hybrid teams, in particular, because it supports team maturity and reduces conflicts among individuals, which helps the team to focus on overcoming challenges. We recommend that teams include training on equity, diversity, and inclusion in their onboarding processes while creating a safe and welcoming space that respects individual differences. In addition, we recommend the development of activities that go beyond the work and help teammates to establish links, such as informal chats, games, recreation, etc.
\end{itemize}

For academia, the main contribution of this research is the first theory of resilience in software engineering. In addition, the discussions provided in this paper open up several opportunities for studies that can be conducted based on the proposed theory. We demonstrated that resilience in software engineering is an important topic because (1) it has not been studied as much as other factors, such as communication or coordination; (2) resilience is, by definition, the thing teams need to overcome crises; (3) recent crises have prompted a huge shift to hybrid work which, if poorly-configured, sabotages resilience. From this point forward, we expect that the themes and concepts described in this study can be explored by software engineering researchers towards the improvement of hybrid work models.

\subsection{Limitations}
In this section we evaluate our study using Charmaz’s~\cite{charmaz2014constructing} criteria: credibility, originality, resonance, and usefulness. To improve credibility, we provided evidence from direct quotations to findings in Table 2. We also interviewed participants from diverse backgrounds (e.g. gender, race, sexual orientation) to ensure the range and depth of our data. To maximize resonance, we constantly compare our categories with our data. We also did member-checking to ensure our data are aligned with our participants’ experiences. To demonstrate usefulness, we provide practical recommendations for software engineers and organizations to increase work team resilience in a hybrid environment. Finally, for originality, our study extends current theories on resilience to the resilience of hybrid teams, which will be a common work practice in the future years. Generalizing from a sample to a population is not a goal of grounded theory methodology~\cite{carminati2018generalizability}; instead, findings should be \textit{transferable} to other contexts. 

However, this study is not without limitations. First, the study is based on the experiences of a limited number of software professionals. Second, despite our efforts in ensuring diversity when recruiting the participants, only one participant has a disability. Our participants are also mainly located in Brazil and work for US companies or US-based clients, which may limits cross-cultural analysis. Based on our understanding of differences across cultures and organizational behaviours, the theory makes sense in different contexts, although some specific concepts may differ because while some details may differ, the concepts of resilience, maturity, and others make sense across many cultures. 

\subsection{Future Work}
Future research on hybrid configurations is required to understand how the software teams can optimize the positive effects of hybrid work in software development without the risk of disruptions. Further, since grounded theory research does not support statistical generalization from a sample to a population, we plan to create and validate a questionnaire survey~\cite{pfleeger2001principles,wagner2020challenges} to test the extent to which the theory generalizes to a larger and more diverse sample of professionals, in particular from different countries.

In addition, hybrid work in software engineering shed some light on the importance of supporting professionals from underrepresented groups in the software industry (e.g., caregivers (especially mothers), LGBTQIA+ individuals (especially transgender people), and people with disabilities). We cannot afford to have hybrid software teams disintegrating or lower their productivity, as it might risk the job opportunities that currently are more accessible for software professionals from equity-deserving groups. We raised some discussions on this topic in this study; however, there are several aspects related to diversity in software development that we plan to investigate in the near future. 

\section{Conclusion}
In summary, we conducted a constructivist grounded theory study by interviewing software professionals about their experience working in hybrid environments. Resilience emerged from our analysis as a core category. The main contribution of this study is the resulting theory of resilience in hybrid software engineering (Fig.~\ref{fig:theory}). 

Team resilience is crucial in hybrid software teams due to the constant change of the environment and its structures, which requires teams to quickly adapt to the dynamic environment. Adopting a hybrid team structure can undermine team resilience by hindering maturation, which presents an important puzzle for researchers. Our theory presents five factors that influence the resilience of hybrid teams, namely, belonging, conflict management, proactiveness, organizational support and diversity. The first three factors, along with communication and coordination effectiveness influence resilience by promoting group maturity. Diversity can increase conflicts, which can be moderated by strong team belonging; however it also helps teams to be more creative in dealing with problems.

Finally, as software professionals strongly value hybrid work, this model is expected to be more recurrent in software organizations, in particular, to avoid the turnover of highly skilled professionals. Therefore, team resilience is essential in the software industry nowadays. 

\section*{Data Availability} \label{sec:DataAvailability}
Example interview guides, an extended chain of evidence table, and ethics approval are available at \url{https://bit.ly/postpandemicresilience}.

\section*{Acknowledgments}
The authors would like to thank all of the participants who took part in this study. This project was supported by NSERC Discovery Grant RGPIN-2020-05001, Discovery Accelerator Supplement RGPAS-2020-00081 and the MITACS Globalink program.

\balance

\ifCLASSOPTIONcaptionsoff
  \newpage
\fi

\bibliographystyle{IEEEtran}
\bibliography{bib.bib}

\end{document}